\documentclass[fleqn,11pt,a4paper]{article}
\usepackage[dvips]{epsfig}

\newcommand{\lsim}{\rlap{\raise 2pt \hbox{$<$}}{\lower 2pt \hbox{$\sim$}}}
\newcommand{\gsim}{\rlap{\raise 2pt \hbox{$>$}}{\lower 2pt \hbox{$\sim$}}}

\newcommand{\etal}{{\it et al.}}
\newcommand{\ie}{{\it i.e.\ }}
\renewcommand{\d}{\mbox{\rm d}}
\newcommand{\eg}{{\it e.g.\ }}

\def\omit#1{\relax}

\newcommand{\ffig}[5]{\begin{figure}[#1]\vfill\begin{center}
\mbox{\epsfig{figure=#2,width=#3}}\caption{#4}\label{#5}
\end{center}\vfill\end{figure}}
             
\begin{document}
\thispagestyle{empty}   
\noindent
TSL/ISV-96-0136  \hfill ISSN 0284-2769\\
March 1996       \hfill                \\
\\
\vspace*{5mm}
\begin{center}
  \begin{LARGE}
  \begin{bf}
Particle Production in the Interstellar Medium\\
  \end{bf}
  \end{LARGE}
  \vspace{5mm}
  \begin{Large}
G.~Ingelman$^{a,b,}$\footnote{ingelman@tsl.uu.se} and M.
Thunman$^{a,}$\footnote{thunman@tsl.uu.se} \\
  \end{Large}
  \vspace{3mm}
$^a$ Dept. of Radiation Sciences, Uppsala University,
Box 535, S-751 21 Uppsala, Sweden\\
$^b$ Deutsches Elektronen-Synchrotron DESY,
Notkestrasse 85, D-22603 Hamburg, Germany\\
  \vspace{5mm}
\end{center}
\begin{quotation}
\noindent
{\bf Abstract:}
The flux of neutrinos and photons originating from cosmic ray
interactions with the interstellar medium in the galaxy is calculated
based on current  models for high energy particle interactions.  The
contribution from a possible dark matter halo of the galaxy is
considered.  The photon flux gets a non-trivial attenuation due to
interactions with the  cosmic background radiation. The neutrino fluxes
are compared with those originating from the Earth's atmosphere as well
as from active galactic nuclei.   
\end{quotation}

\section{Introduction}

Cosmic rays of galactic and extra-galactic origin will interact in high
energy collisions with the interstellar medium  of our galaxy and
produce secondary particles \cite{Domokos93,Berezinsky93,DePaolis95}. 
These are mainly mesons that decay and give rise to a flux of
muons, neutrinos and photons. The very low density of the interstellar
medium imply that the interaction lengths of the secondary particles is
long compared to their decay length, such that the mesons will decay
before loosing energy in secondary interactions. This is also the case
for the muons which will decay giving neutrinos. This is the
opposite of the situation for cosmic ray particles interacting in the
Earth's atmosphere \cite{GIT}, where meson typically loose energy in
interactions before decaying. The fluxes of high energy neutrinos and
photons from the interstellar medium could therefore be larger than
those from the atmosphere, although the initial production rate of
mesons is smaller.

A measurement of these fluxes could potentially give valuable
information  about the distribution of matter and cosmic rays in the
galaxy, which  could be of great importance in determining the origin
of the cosmic rays.  Understanding the flux from the disc of the Milky
Way could also be the  starting point in a search for baryonic dark
matter in a spherical halo  around the Milky Way
\cite{Domokos93,DePaolis95}.

In addition, these fluxes from the interstellar medium constitute a 
background in searches for other, more spectacular cosmic sources and
must therefore be known in order to extract the desired signal. For
example, there is much interest in neutrinos from Active Galactic
Nuclei (AGN) (see for example \cite{Stecker95} and references therein). 

In this paper we present a realistic calculation of the neutrino and
photon fluxes to be expected from cosmic ray particle interactions in
the interstellar medium. These fluxes are derived from complete events 
obtained by detailed Monte Carlo simulations based on state-of-the-art 
models for high energy particle collisions.  In section 2 we present
the model including the cosmic ray energy  spectrum, the interstellar
matter distribution in the Milky Way and the model for particle
production in high energy elementary particle collisions. The resulting
fluxes are presented in section 3, where also the attenuation of the
photon flux due to interactions with the cosmic background radiation is
demonstrated to give a significant effect. Section 4 considers the
mentioned dark matter halo and the fluxes that would arise from it. We
end, in section 5, with a discussion and comparison with estimated
neutrino fluxes from active galactic nuclei.  

\section{The calculational method}

The fluxes of neutrinos and gammas that could be observed at the Earth
is basically given by the production rate in the galaxy and the 
probability that the produced particle reaches the Earth. The fluxes
can be expressed as an integral along the line of sight from the Earth
($r=0$) to the edge of the galaxy ($r=R$) 
\begin{equation}
\label{eq:production}
\frac{\d \phi_{\nu/\gamma}(E)}{\d E} = \int^{R}_{0}\d r
\int^{\infty}_{E} \d E' \phi_{p}(E',r)\, \rho(r)\, \frac{\d
\sigma_{\nu/\gamma}(E',E)}{\d E}\, A(E,r)\, P(r),
\end{equation}
where the primary cosmic ray flux $\phi_{p}(E',r)$ is folded with the 
interstellar matter density $\rho(r)$ and the differential cross
section for producing neutrinos or photons. An attenuation factor
$A(E,r)$ is also included to account for the loss of flux due to
interactions with the interstellar medium and $P(r)$ is the probability
that a particle is directed toward the Earth. $A(E,r)$ is calculated
separately and folded with the non-attenuated fluxes. Its effect 
is important for the gamma flux were interactions with the
cosmic background radiation producing electron-positron pairs occur. 

The integration over $r$ can be greatly simplified since one may
calculate  the flux from a `unit' column density and then scale with
the appropriate  integrated density in any given direction. This
procedure is possible since the distance between a typical production
point $r$ and the Earth is large  compared to the decay length for all
particles.  The energy integral can be obtained based on Monte Carlo
simulations of  high energy particle collisions giving complete final
states including  particle decays. For this purpose we have employed
the   Lund Monte Carlo programs \cite{pythia}. 

\subsection{The cosmic rays}

The flux of cosmic rays in the Milky Way is not well known. The flux
measured at the Earth is found to be isotropic to a high degree, the
anisotropy being $\lsim 5\%$ \cite{Gregory82}. This isotropy can arise
in two ways. First, the bulk of the flux is of extra-galactic origin
with a small component from processes in the Milky Way, \ie the flux is
universal. Secondly, it is of galactic origin with only a component at
the highest energies of extra-galactic origin. The flux at lower energy
is in this case captured by the magnetic field of the galaxy and
thereby appearing to be isotropic at the Earth. A flux of
extra-galactic origin will not be significantly attenuated when passing
through the galaxy, since only a few per mill of the cosmic ray
particles will interact. This attenuation effect is smaller than other
uncertainties concerning the primary flux, and the flux can be treated
as homogeneous. 

Assuming that the Earth is not at a unique site in the galaxy,  it
follows that the flux is the same everywhere in the galaxy. Therefore,
we assume the flux observed at the Earth to be uniformly and
isotropically  distributed throughout the galaxy.  The energy
dependence of the flux can then be parameterized as 
\cite{Gaisser90,Volkova87}
\begin{equation}
\label{eq:initflux}
\phi_{N}(E)\left[\frac{\mbox{nucleons}}
{\mbox{cm$^{2}$\,s\,sr\,GeV/$A$}} \right] = \left\{ \begin{array}{ll}
    1.7\,E^{-2.7} &  E<5\cdot10^6\,\mbox{GeV} \\
      & \\
      174\,E^{-3} & 
      E>5\cdot10^6\,\mbox{GeV}
      \end{array}
      \right.
\end{equation}
The normalisation is here derived \cite{Pal92} from the directly 
measured primary spectrum using balloon-borne emulsion stacks in
JACEE \cite{JACEE}. It agrees (within some 10\%) with more indirectly
derived spectra based on measured atmospheric muon fluxes \cite{MACRO},
and is also compatible with the data discussed in ref.~\cite{Honda95}.
The cosmic ray composition is dominated by protons with only a
smaller   component of nuclei \cite{Pal92,Honda95}. 

\subsection{The model of the Milky Way}

Given the uncertainties concerning the matter distribution in the
galaxy, one can only make a fairly simple model for it. However, this
should also  be adequate for our purpose to estimate the fluxes. Such a
model has been  used by Domokos \etal\ \cite{Domokos93}, but it seems
to overestimate the  column density for high galactic latitudes.
Therefore, a model similar to  the one in \cite{Berezinsky93} will be
used outside the plane of the Milky Way, where the density decreases
exponentially with the distance to the galactic  plane. 

We take the galaxy to be rotationally symmetric in the plane with a
radius of $12\, kpc$ and with a constant constant density of $1\,
nucleon/cm^3$ in the plane. Outside the plane, the density decreases as
\begin{equation}
\rho(h) = \rho_0\,e^{-h/h_0},
\end{equation}
where $h$ is the height above (or below) the disc and $h_0=0.26\,kpc$
is the scale height at the galacto-centric distance of the Earth. The
Earth is in the galactic plane at a distance of $R_{GC}=8.5\,kpc$ from
the centre. 

Stars are not taken into account in this study, although they affect
the fluxes in two ways. First, the cosmic ray particles can interact in
the stellar material producing mesons. These will, however, often have
secondary interactions due to the relatively higher density environment
in the stars. They will thus typically be absorbed or degraded in
energy before they decay and therefore relatively less important
\cite{Domokos93}. Furthermore, the neutrinos and gammas from these
meson decays must traverse the star such that photons will be absorbed
and also the neutrino flux be significantly attenuated. These aspects
are treated in our study \cite{IT-sun} of the neutrino fluxes arising
from cosmic ray interactions in the Sun. Secondly, neutrino and gamma
fluxes from the interstellar medium will be attenuated because of
absorption in stars. However, this is a very small effect since the
solid angle covered by stars is only $\Delta\Omega/\Omega\lsim
10^{-10}\,$ (derived from \cite{Domokos93}).

\subsection{The model for particle production}

To specify the energy spectra of secondaries in cosmic ray collisions
with  the nuclei of the interstellar medium a model for particle
production is needed. With the cosmic rays being predominantly protons
and the interstellar matter mainly  hydrogen, and only some smaller
fraction of heavier nuclei in both cases, the interactions producing
the secondary particle fluxes are dominantly proton-proton collisions.
Contributions of nuclear collisions can also be treated as
nucleon-nucleon interactions, since the nuclear binding energies are
negligible and other nuclear effects  have only little influence on the
high energy secondary particles that are  of interest to us. Therefore,
the primary interactions can be taken as  proton-proton collisions. 

The dominating source of neutrinos and gammas are then the decays of
the light mesons ($\pi$ and $K$) and muons plus a small contribution
from other  heavier hadrons.  Given the low density of the interstellar
medium, secondary interactions are  rare and will essentially not
happen before unstable particles decay. Therefore, with all light
hadrons decaying, the contribution from charmed and heavier particles
will not be important as opposed to the case of interactions in the
Earth's atmosphere \cite{GIT}. The dominance of light hadrons (pions) 
will only disappear above their critical energy 
($\varepsilon_{\pi}^{critical}\sim 10^{21}\,GeV$ \cite{Domokos93}),  
where their decay becomes less probable compared to their interaction. 
Heavy quark production is therefore not explicitly considered in the
simulation, but their contribution is  estimated based on the
spectrum-weighted moment method \cite{GIT} and shown to be negligible. 

The production of light hadrons (not containing heavy quarks)  is
dominantly through minimum bias hadron-hadron collisions. The strong
interaction mechanism is here of a soft non-perturbative nature that
cannot be theoretically calculated from first principles, but must be
modelled. In the successful Lund model \cite{lund} hadron production
arise through the fragmentation of colour string fields between partons
scattered in semi-soft QCD interactions \cite{pythia}. The essentially
one-dimensional colour field arising between separated colour charges
is described by a one-dimensional flux tube whose dynamics is taken as
that of a massless relativistic string. Quark-antiquark pairs are
produced from the energy in the field through a quantum mechanical
tunneling process. The string is thereby broken into smaller pieces
with these new colour charges as end-points and, as the process is
iterated, hadrons are formed. These obtain limited momenta transverse
to the string (given by a Gaussian of a few hundred MeV width) but
their longitudinal momentum may be large since it is given by a
probability function in the fraction of the available energy-momentum
in the string system taken by the hadron. All mesons and  baryons in
the basic multiplets may be produced and the subsequent decays are
fully included. The iterative and stochastic nature of the process is
the basis for  the implementation of the model in the {\sc Jetset}
program \cite{pythia}. The parameters in the program were taken at
their default values, except that  also long-lived particles had to be
treated as unstable and their decay  simulated. 

A non-negligible contribution to the inclusive cross section is given
by diffractive interactions.  These are also modeled in {\sc Pythia}
\cite{pythia} using cross sections from a well functioning Regge-based
approach and  simulating the diffractively produced final state using
an adaptation of  the Lund string model. These diffractive events are
included in our  simulations and contribute rather less than 10\% to
the final results.  

\section{Results}

As mentioned, the production rate is only a function of the density
since the cosmic ray flux is assumed to be position independent within
the galaxy. The energy integral of Eq.\,(\ref{eq:production}) has been
calculated for a `unit' column density of $1\,GeV/cm^3$ over a distance
of $1\,kpc$ ($\sim5\,mg/cm^2$). This was performed by choosing energies
between $10^2$ and $10^{10}\,GeV$ from a flat distribution in
$log_{10}E$, and assigning a weight to account for the detailed form of
the primary cosmic ray flux in Eq.~(\ref{eq:initflux}). Proton-proton
collisions at these energies were then simulated using {\sc Pythia}
\cite{pythia} resulting in complete events. After all particle decays,
neutrinos and photons were recorded and filled in energy-histograms
applying the weight for the event.  

\subsection{The neutrino fluxes}

The resulting fluxes of muon and electron neutrinos are shown in 
Fig.\ref{fig:nuflux}. For muon neutrinos the dominating source is muon
and pion decay, while for electron neutrinos muon decay dominate
strongly with only a few per cent from kaon decay and a totally
negligible fraction from neutron decay. 

This is rather different from the other sources of cosmic ray induced
neutrinos in the vicinity of the Earth, \ie the Sun's and the Earth's
atmosphere. In case of the Earth's atmosphere the distance involved is
so short that the muon can be considered almost stable. The most
important sources of muon (electron) neutrinos are therefore the decays
of pions and kaons (kaons), with charmed particles being the dominating
source at very high energies \cite{GIT}. A comparison with these
atmospheric fluxes is made below in section~5 (Fig.\,6). For the Sun,
muons cannot be considered stable, and give a significant contribution
to the fluxes. The muons, however, loses energy through electromagnetic
interactions and are therefore less important at higher energies
compared with the light mesons \cite{IT-sun}.

Because of the small cross section for $\nu$-hadron and $\nu-\gamma$
interactions, the $\nu$-fluxes will not be noticeably attenuated due to
interactions with the interstellar medium or the cosmic microwave
background. The attenuation due to interactions with stellar material
is also negligible since, as mentioned, the stars cover only a very
small solid angle.

\ffig{tb}{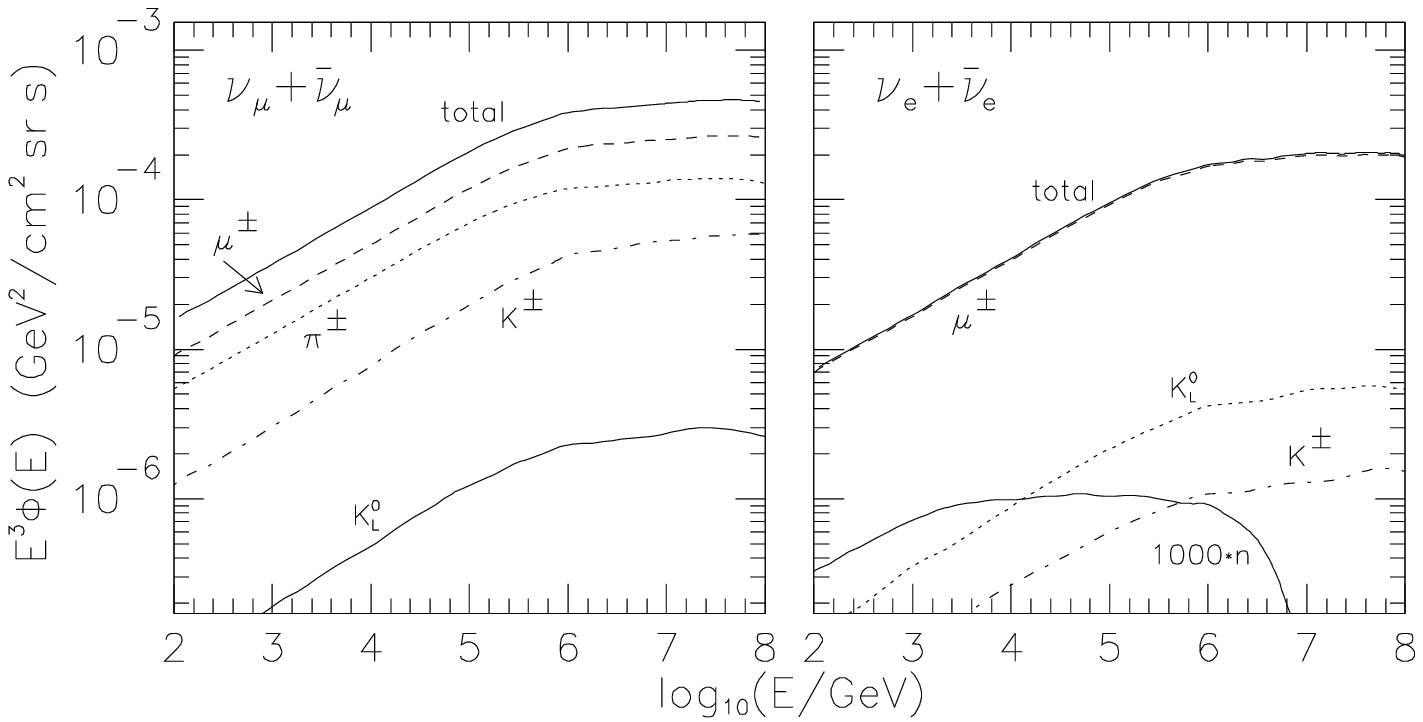}{14cm}{\it 
The $E^3$-weighted flux of muon and electron neutrinos from interactions
of cosmic rays with the interstellar medium in a column of length
1\,kpc and density $1\,nucleon/cm^3$. The total flux is shown (full
curve) as well as the contributions from the decay of the indicated
particles.}{fig:nuflux}

\subsection{The photon flux}

As opposed to the neutrino fluxes, the photon or gamma flux is
attenuated by interactions. Photons of very high energy can interact
with the cosmic  microwave background to produce particles. Unless the
energy is extremely  high, one need only consider electron-positron
production, \ie  $\gamma \gamma \to e^+e^-$. 

The cosmic microwave background is isotropic and has an energy
spectrum  given by 
\begin{equation}\label{eq:CMB}
\frac{dn(E)}{dE}=\frac{1}{\pi^2(\hbar c)^3}~\frac{E^2}{exp(E/kT)-1} +
\Theta (10^{-5}-E)\, \frac{3.3\cdot 10^7}{E},
\end{equation}
with $E$ in $eV$. The first part is the thermal background
spectrum at temperature $T$ and the second part is a very simple
parameterisation of the radio-wave background at very low energy
\cite{Kolb}. The latter is far from as accurately know as the former,
but since it only influences the flux above $\sim 10^7\,GeV$ it is not
of significant importance for our purposes.
 
To calculate the attenuation one needs to account for both the
energy-spectrum of the background radiation and the angle between the
two photons. We do this by defining an
effective thickness ($mbarn/cm^3$) given by
\begin{equation}
<\sigma n>(E_{\gamma}) = \frac{1}{4 \pi} \int_{0}^{2 \pi} d\phi 
\int_{-1}^{1} d(cos\theta) \int_{E_{th}(\theta)}^{\infty} dE 
\frac{dn(E)}{dE} \sigma^{\gamma\gamma \rightarrow \ell\overline{\ell}}
(2E_{\gamma}E(1+cos\theta)),
\label{eq:nsigma}
\end{equation}
where $\phi$ and $\theta$ are the azimuthal and polar angles between
the interacting gammas in the galactic `laboratory system'.
$\frac{dn(E)}{dE}$ is the cosmic background radiation in
Eq.~(\ref{eq:CMB}) and  $\sigma^{\gamma\gamma \rightarrow
\ell\overline{\ell}}(s)$  is the cross section for charged lepton pair
production as a function of the  CM-energy $\sqrt{s}$, which we have
calculated. It is
sufficient in this context to use the leading order formula for this 
cross section and to consider $e^+e^-$production only. The threshold
energy for a microwave background photon is 
$E_{th}(\theta)=2m_{\ell}^2/E_{\gamma}(1+cos(\theta))$ in terms of the
energy $E_{\gamma}$  of the incoming photon.

\ffig{bt}{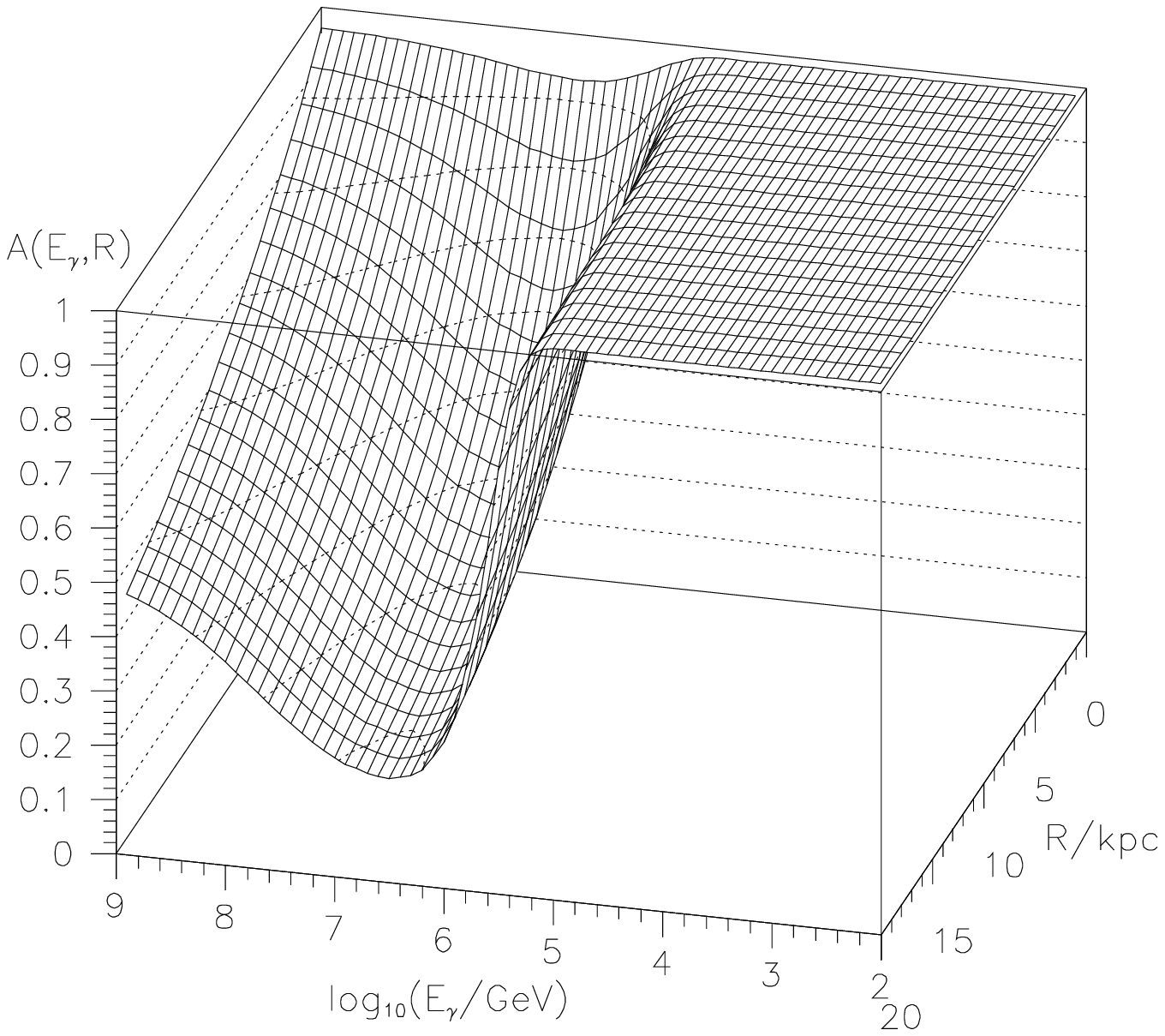}{9cm}{\it 
The attenuation factor, Eq.~(\protect\ref{eq:supgamma}), for the photon
flux as a function of the photon energy $E_{\gamma}$ and the distance
$R$ from the  Earth to the edge of the galaxy.} {fig:supgamma}

Since the production of photons is isotropic along the line of sight
within the plane of the Milky Way an attenuation factor can be derived
and applied to the flux obtained from the Monte Carlo simulation. The
attenuation factor is obtained by integrating over the line of sight
and weighting each production point with the probability
($e^{-r\,<\sigma n>}$) that a produced photon reaches the Earth, 
\begin{equation}
A(E_{\gamma},R)=\frac{1}{R<\sigma n>(E_{\gamma})} (1-e^{-R<\sigma
n>(E_{\gamma})})
\label{eq:supgamma}
\end{equation}
where $R$ is distance to the edge of the Milky Way. A numerical
evaluation of this factor is shown in Fig.~\ref{fig:supgamma}. The dip
structure in the attenuation factor is a consequence of folding two
peaked functions. The cosmic microwave background radiation which
peaks at $T\sim7\,K$, and the electron-positron production cross
section peaked at $s\sim13\,m_e^2$.

With this attenuation factor, we then obtain the attenuated photon flux
shown in Fig.\,\ref{fig:gammaflux} and compared to the unattenuated
flux; both are in the direction towards the centre of the Milky Way.
The attenuation dip structure is obviously a reflection of the
structure in Fig.\,2. Another such dip structure would also appear at
an energy of $(m_{\mu}/m_e)^2\,10^{6.5}\,GeV\,\approx\,10^{11}\,GeV$,
when the reaction $\gamma \gamma \to \mu^+ \mu^-$ becomes energetically
possible. For such high energies is, however, our parameterisation of
the primary flux inadequate and our method therefore no longer
applicable. 

\ffig{tb}{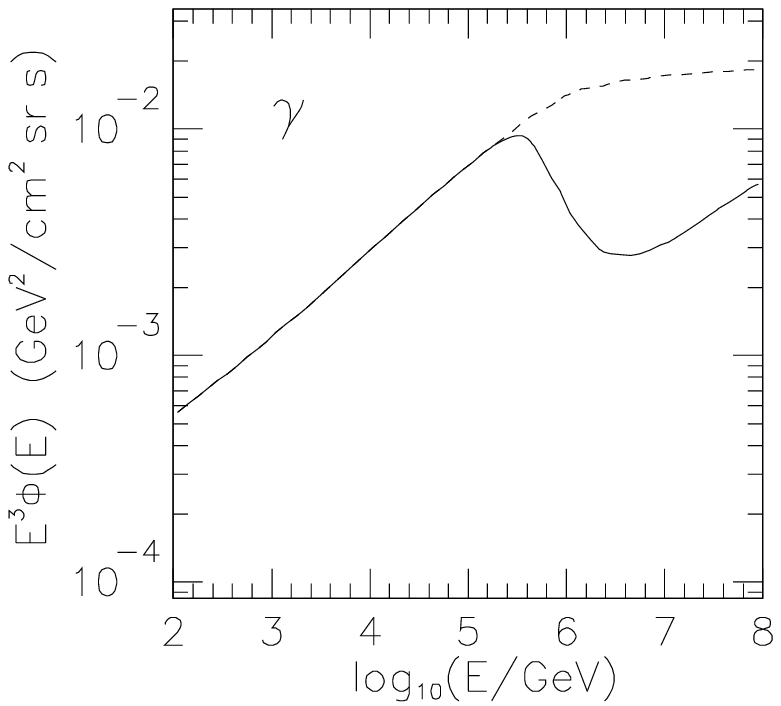}{7cm}{\it 
The attenuated (solid line) and unattenuated (dashed line) 
$E^3$-weighted flux of photons from interactions of cosmic rays with the 
interstellar medium of density $1\, nucleon/cm^3$ integrated over the 
whole Milky Way in the direction to its center. 
} {fig:gammaflux}

\subsection{Parameterisation of the fluxes}

The resulting fluxes can be parameterised in a form inspired by the
analytic formalism used by Berezinsky \etal\ \cite{Berezinsky93}. The
photon flux, which is attenuated through pair production as discussed,
is better treated in two steps; one for the production and one for the
attenuation. The non-attenuated fluxes are parameterised as
\begin{equation}
\label{eq:fitflux}
\phi(E)=\left\{\begin{array}{ll}
     R\delta\,N_{0}\,E^{-\gamma-1} & E<E_{0} \\
      & \\
     R\delta\,N_{0}'\,E^{-\gamma'-1} & E>E_{0} \\
     \end{array}
\right.
\end{equation}
with the fitted parameter values given in Table~\ref{tab:fitflux}.
($N_{0}'$ is not fitted, but given by the continuity condition at
$E_{0}$.) $R\delta$ is the column number density, $\delta$ is the
average number density in $nucleons/cm^3$ and $R$ is the distance to
the edge of the galaxy in $kpc$.  By inserting the corresponding values
from our model, or another model,  of the galaxy one can obtain the
fluxes from this parameterisation.  Results are shown in
section~\ref{sec:discussion} and discussed in connections with other 
sources. 

The attenuation factor $A(E,R)$ can be obtained from
Eq.~(\ref{eq:supgamma}) using the effective thickness parameterised as 
\begin{equation}
\log \left\{ <n\sigma>(E_0)\right\} = 
     -25.34+6.82\,\sqrt{\log E_0-4.99}\,e^{-0.29\,(\log E_0-4.99)}\,
 \hspace{6mm} E_0>10^5\,GeV
\end{equation}
whereas for $E_0<10^5\,$GeV the pair production cross section
negligible small and the attenuation factor effectively unity, see
Fig.~\ref{fig:supgamma}.

\begin{table}
\begin{center}
\begin{tabular}{|l||l|l|l|l|l|}  \hline
particle & \multicolumn{}{c}{$N_{0}$} & \multicolumn{}{c}{$\gamma$} & 
\multicolumn{}{c}{$E_{0}$} & \multicolumn{}{c}{$\gamma'$} & 
\multicolumn{}{c}{$N_{0}'$} \\ \hline
$\nu_{\mu}+\bar{\nu_{\mu}}$ & $3.0\cdot 10^{-6}$ & 1.63 & $4.7\cdot 10^{5}$ 
& 1.95 & $ 1.9\cdot 10^{-4}$ \\ \hline
$\nu_{e}+\bar{\nu_{e}}$     & $1.3\cdot 10^{-6}$ & 1.63 & $4.2\cdot 10^{5}$ 
& 1.95 & $ 8.3\cdot 10^{-5}$ \\ \hline
$\gamma$                    & $4.7\cdot 10^{-6}$ & 1.63 & $7.6\cdot 10^{5}$ 
& 1.94 & $ 3.1\cdot 10^{-4}$ \\ \hline
\end{tabular}
\end{center}
\caption{\em 
Values of the parameters in Eq.\,(\protect\ref{eq:fitflux}) obtained
from fits to the neutrino fluxes in Fig.\,\protect\ref{fig:nuflux}  and
the unattenuated photon flux in Fig.\,\protect\ref{fig:gammaflux}.}
\label{tab:fitflux} \end{table}

\section{Fluxes from a possible dark matter galactic halo}

Cosmic rays could also interact with matter in a galactic halo and, 
depending on the properties of this matter, give a neutrino flux of 
interest.  If the halo matter consists of weakly interacting massive
particles (WIMPS)  the interaction cross section is very small and
would give a negligible  neutrino flux.  However, with dark matter of
hadronic nature this source could give  a measurable flux. In the
scenario of massive cold halo objects (MACHO's), the interactions would
take place in objects of high density. The interaction  lengths of the
produced secondary mesons would then be shorter than their  decay
lengths such that they would loose energy or be absorbed before
decaying  into leptons. This would result in a lepton flux at very low
energies only,  which at the Earth would have to compete with the high
atmospheric flux.  Recent measurements by the MACHO collaboration
\cite{MACHO95} show, however, that these objects can only make up
$\sim20$\% of the dark matter. This leaves the possibility for a
gaseous hadronic component of the dark matter as  discussed by
De\,Paolis \etal\ \cite{DePaolis95}.  They calculate gamma ray fluxes
in the energy range $1-10^6\,GeV$ from decays of $\pi^0$'s produced in
cosmic ray interaction.

\ffig{tbh}{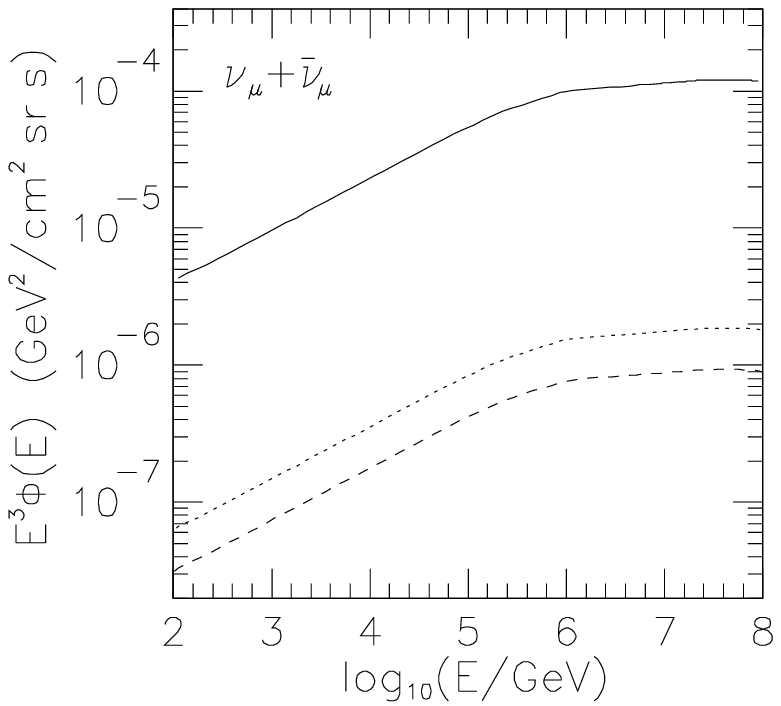}{7cm}{\it The
$E^3$-weighted flux of muon neutrinos from a possible dark matter
galactic  halo induced by interactions of cosmic rays of galactic 
(dashed line) and extra-galactic (dotted line) origin. Shown for
comparison is also the flux from the normal interstellar medium
(section 3.1) at high galactic latitudes (solid line).} {fig:darkhalo}

Following the approach of De\,Paolis \etal\ \cite{DePaolis95} we have
calculated the corresponding neutrino fluxes. The gaseous hadronic
matter distribution in the halo is assumed to be of the form 
\begin{equation} \label{halodensity}
\rho(R) = \rho_0\,f\,\frac{a^2+R_{GC}^2}{a^2+R^2}, 
\end{equation} 
where $\rho_0\approx0.3\,nucleons/cm^3$ is the local dark matter
density and $f\sim0.5$ is the fraction of dark matter in form of
gaseous nucleons. $R$ is the distance from the galactic centre to an
arbitrary point and $R_{GC}\sim8.5\,kpc$ to the Earth, whereas
$a\sim0.5\,kpc$ is  the galactic core radius. 

The flux of cosmic rays in the halo may consist of two parts.  First,
particles that escapes from the disc, which probably give a flux that 
decreases in the same way as the halo density does, \ie
Eq.\,(\ref{halodensity}), unless it is confined by some unknown
mechanism. Secondly, there may be a flux of extra-galactic origin,
which can be assumed to be isotropic. De\,Paolis \etal\
\cite{DePaolis95} do not consider energies higher than $10^6\,GeV$ and
are therefore dominated by the first part. The normalisation can then
be estimated from theoretical arguments \cite{Breitschwerdt91}
resulting in $\sim 1/500$ times that at the Earth.  Thus we have the
cosmic ray flux in the halo 
\begin{equation}\label{haloflux}
\phi_p(E,R) = \frac{\phi_{p,\oplus}(E)}{500}\,\frac{a^2+R_{GC}^2}{a^2+R^2}.
\end{equation}
Part of this flux is directed away from the galaxy such that the
secondary  flux will not reach the Earth. This could be accounted for
by an efficiency  factor, but since it is unknown it is not taken into
account. 

The fluxes of neutrinos are then obtained by folding the cosmic ray
flux and the halo density and integrating the production along the line
of sight analogously to the galactic flux (Eq.\,(\ref{eq:production})).
The resulting flux is shown by the dashed curve in 
Fig.\,\ref{fig:darkhalo} and seen to be significantly smaller than  the
galactic flux at high latitudes (where it is lowest). One should
remember that this flux from the halo is a crude  estimate and may be
seen as an upper limit due to the mentioned  neglect of the efficiency
factor.

Extra-galactic cosmic rays are also believed to contribute to the flux
in the halo, especially at very high energies. The normalisation of
this component is unknown, but could be as high as given by the high
energy flux at the Earth. In order to compare with the flux  induced by
cosmic rays escaping from the galaxy, we take the normalisation to be
same as in Eq.\,(\ref{haloflux}),
\begin{equation} \label{haloflux2}
\phi_p(E,R) = \frac{\phi_{p,\oplus}(E)}{500}.
\end{equation}
The neutrino flux is here obtained in the same way as above, and shown
by the dotted curve in  Fig.\,\ref{fig:darkhalo}.

Although it seems that the flux from the halo cannot compete with that
from the galactic disc, it is still conceivable that its magnitude is
significantly higher than here estimated. This would happen if the
primary cosmic rays are completely of extra-galactic origin and
therefore the same as that at the  Earth, \ie  Eq.\,(\ref{haloflux2}) 
reduces to $\phi_p(E,R) = \phi_{p,\oplus}(E)$. This would give a
neutrino flux that is about a factor five higher than the corresponding
interstellar flux. Although, such high extra-galactic flux is not 
probable, a higher flux than in Eq.\,(\ref{haloflux2}) is not
unrealistic. 

From the above discussion, the prospects for detection of a dark halo
based on neutrino flux measurements does not look promising. In
particular, this flux is considerably lower than the atmospheric flux
\cite{GIT} and the estimated one from active galactic nuclei, as
discussed in section~5 (Fig.\,6).

\section{Concluding discussion} \label{sec:discussion}

The neutrino fluxes shown above in Fig.\,\ref{fig:nuflux}  are given
for our basic `unit column'  of interstellar matter. Using these one
may obtain the integrated  flux from any direction. In
Fig.\,\ref{fig:compflux2} we show our resulting muon neutrino flux
(full curves) from the direction towards  the center of the Milky Way
(highest flux) and orthogonal to the galactic  plane (lowest flux).  In
Fig.\,\ref{fig:compflux2}a  we compare with the original results of two
earlier calculations by   Domokos \etal\ \cite{Domokos93} and by
Berezinsky \etal\ \cite{Berezinsky93}. The differences between these
curves are mainly related to the  different assumptions concerning the
interstellar matter density profile and the normalisation of the
cosmic ray spectrum. Thus, they illustrate the  uncertainty due to
these inputs to the different calculations. 

\ffig{b}{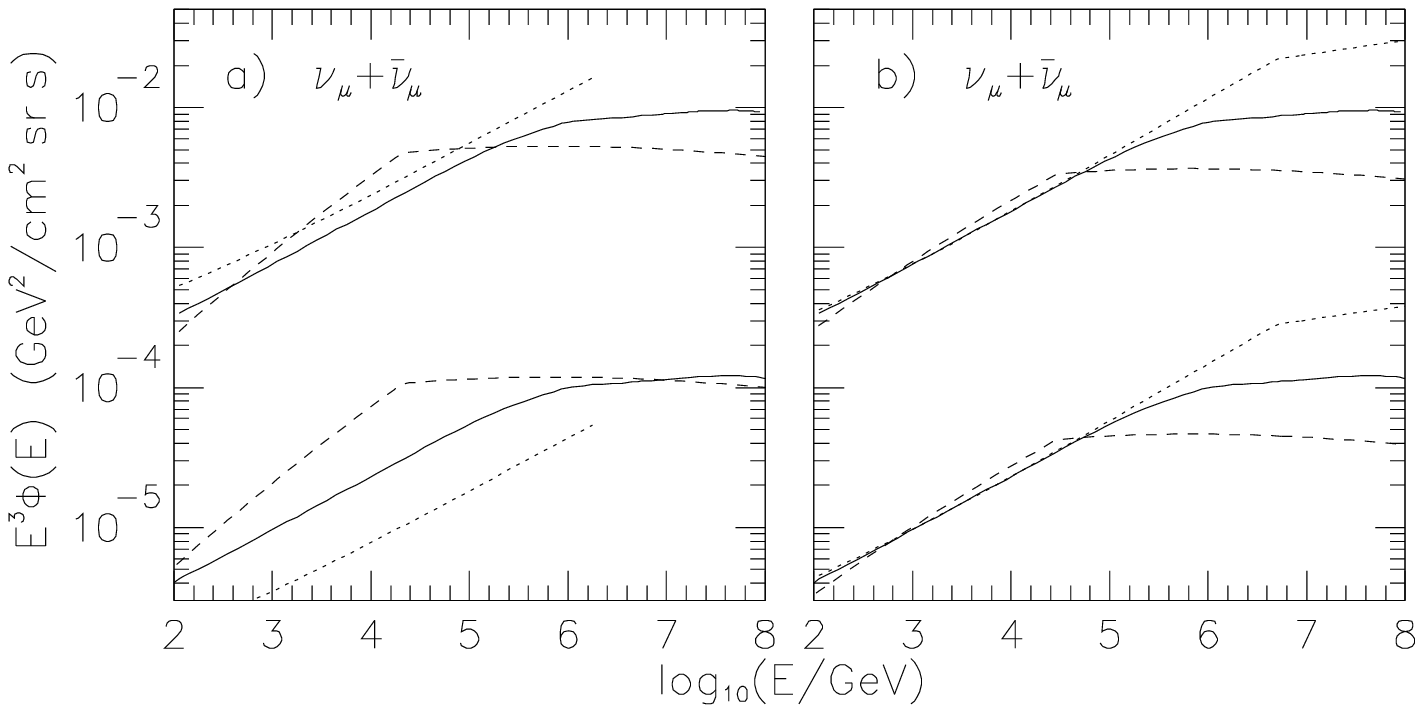}{14cm}
{\it  The $E^3$-weighted fluxes of interstellar muon neutrinos in the
direction  towards the center of the Milky Way (upper set of curves)
and orthogonal to the galactic plane (lower set of curves).  Comparison
of our results (full lines) with those by Domokos \etal 
\protect\cite{Domokos93} (dashed lines) and by Berezinsky \etal
\protect\cite{Berezinsky93} (dotted lines).  In (a) the original
results are used, whereas in (b) the results of 
\protect\cite{Domokos93} and \protect\cite{Berezinsky93} are modified 
to have the same galactic density profile and cosmic ray flux as in
our  calculation. } {fig:compflux2}

In order show other differences between these different calculations we
have  recalculated the results of \cite{Domokos93} and
\cite{Berezinsky93}  using their formalisms but with our density and
cosmic ray parametrisations. The results are shown
Fig.\,\ref{fig:compflux2}b and demonstrate a close agreement for
energies up to $10^5$--$10^6\: GeV$. At higher energies there are,
however, significant differences due to the treatment of the change in
slope of the cosmic ray spectrum, {\it cf.}\ Eq.\,(\ref{eq:initflux}).
In \cite{Berezinsky93} energies above the `knee' are not considered,
and the naive extrapolation to higher energies reults in an
overestimated flux. About a factor three excess at the highest energies
is expected based on  an estimate using the analytic method with
spectrum-weighted $Z$-moments  \cite{GIT,Gaisser90,Lipari}, in
agreement with the effect seen in Fig.\,\ref{fig:compflux2}b. In
\cite{Domokos93} this change of slope is included, but its effect on
secondary particles cannot be fully taken into account since their
calculation is based on an analytic method. One must here make
assumptions about from what average primary energy a given neutrino
comes, and thereby specify how the `knee' from the change in the 
cosmic ray energy spectrum is transported into a `knee' in the final
neutrino spectrum.  This problem does not occur in our Monte Carlo
method, since the neutrino  spectra are here obtained through an
event-by-event simulation taking the  fluctuations into account. Our
results are therefore more reliable in this respect.  

There is also a contribution to the interstellar flux from semileptonic
decays of charmed and heavier hadrons.  We have estimated this based on
the analytic method with spectrum-weighted  moments
\cite{GIT,Gaisser90,Lipari} using our previoulsy calculated 
$Z$-moments for charm particle production and decay \cite{GIT}. Taking
also the contribution from muon decay into account, we find a muon
neutrino flux which is only contributing about $2\cdot10^{-4}$ to the
interstellar flux and a factor two higher contribution  for the
electron neutrino flux (due to a lower interstellar electron neutrino
flux). The smallness of this charm contribution justifies the neglect
of it in our Monte Carlo treatment. 

From Fig.\,\ref{fig:compflux2}, and these considerations, one can also 
conclude that the uncertainties originating from the particle physics
input are not larger than other uncertainties. Our Monte Carlo
calculation also confirmes that the approximation done with the
analytic method are justfied except at the highest energies where
special precautions must be taken.

\ffig{bt}{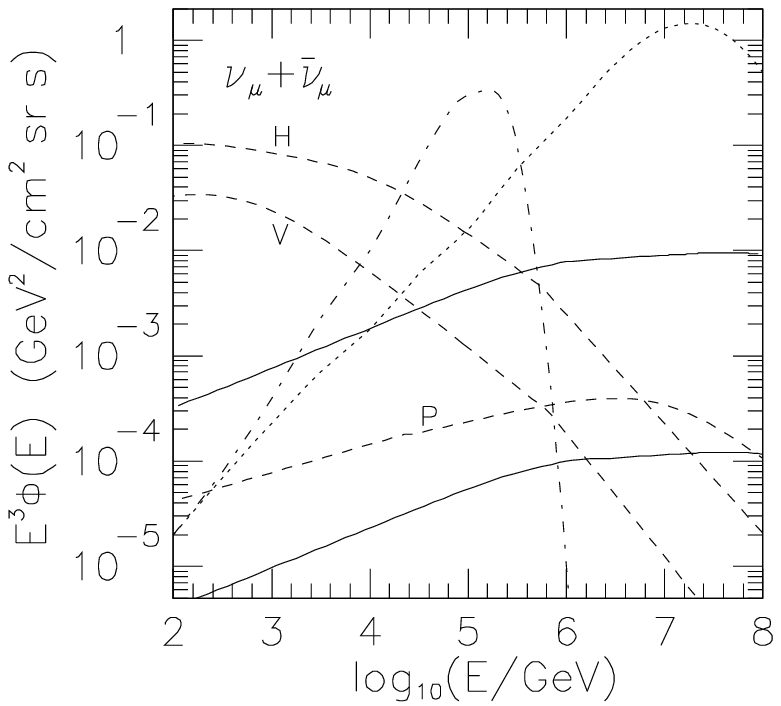}{7cm}{\it 
$E^3$-weighted fluxes of muon neutrinos at the Earth from cosmic ray 
interactions with the interstellar medium looking towards the galactic
centre (upper solid curve) and orthogonal to the galactic plane (lower
solid curve) as calculated in this study, and the flux from the
Earth's atmosphere in the vertical (V) \protect\cite{GIT} and
horizontal (H) \protect\cite{Volkova79} direction originating from
conventional $\pi ,K$-meson decays as well as from prompt (P) charm
decays \protect\cite{GIT}. For comparison, the diffuse flux of muon
neutrinos from  active galactic nuclei as predicted in
\protect\cite{Protheroe} (dotted line) and in \protect\cite{Sikora}
(dash-dotted line) are shown.}{fig:compflux3}

The fluxes from the interstellar medium should be compared with those
from other sources in order to establish their observabelness and their
significance  as a primary interest of study or as a potential
background.  This is done in Fig.\,\ref{fig:compflux3} where our
result on the muon  neutrino flux is compared with those from cosmic
ray interactions in  the Earth's atmosphere and with two predictions of
the diffuse flux from active galactic nuclei. The interstellar flux is
seen to be considerably lower than the atmospheric  flux, except at the
highest energies. It will therefore be very hard to observe with
detectors on the Earth. At the highest energies  considered, the
interstellar flux is of comparable magnitude as the  prompt atmospheric
flux (from charm decays). The latter is almost direction independent up
to $\sim10^7\,GeV$ such that the curve shown  for the vertical
direction applies in essentially the whole energy range; the horizontal
flux being slightly higher for the highest energies only.  Given the
small absolute scale of the fluxes at these energies, the event  rate
will be very low in the neutrino telescopes currently under 
construction. The interstellar flux becomes important compared to the
vertical atmospheric flux only at energies above $10^4\,GeV$. The
corresponding event rate in a detector of $3\cdot10^4\,m^2$, \eg {\sc
Amanda}, would be $\sim0.5/year$ in a cone of opening angle
$10^{\circ}$ directed towards the centre of the Milky way.

Our calculated neutrino flux from the interstellar matter is
significantly lower than the one from active galactic nuclei, as
demonstrated in  Fig.\,\ref{fig:compflux3}, and should therefore not be
a problematic  background for this extra-galactic source.  On the other
hand, the atmospheric neutrino flux will be problematic in  this
context, since it dominates in the lower energy range. At high energies
where the atmospheric flux is lower, the absolute rate is very low. 

Concerning the flux of electron neutrinos the situation is very similar
to the case of muon neutrinos, except that the atmospheric background
is significantly lower \cite{GIT} and the interstellar flux is a
factor $\sim2$ lower than the corresponding muon neutrino flux
(Fig.\,\ref{fig:nuflux}). 
Since the dominating production process in AGN is decays of pions
followed by the decays of the muons, the electron neutrino flux from
AGN's should also be a factor $\sim2$ lower. Thus, the relative fluxes
from these sources should be the same as for  muon neutrinos. There is,
however, no good technique to detect high energy  electron neutrinos,
so the prospects to search for such a signal is worse.

\ffig{tb}{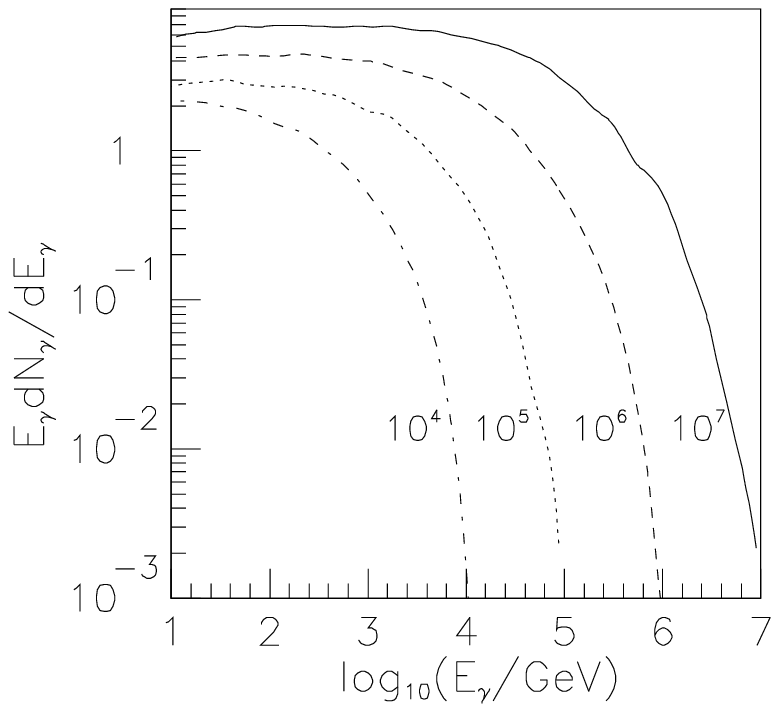}{7cm} {\it 
Energy-weighted spectrum of photons from interactions of cosmic ray 
particles, having the specified energies, with the interstellar matter
(\ie the photon yield). } {fig:gyield}

The yield of photons in cosmic ray interactions with the interstellar
matter has been considered before, \eg  by Berezinsky \etal\ 
\cite{Berezinsky93}. Our corresponding result obtained from the
unattenuated photon flux in section 3.2 is  shown in
Fig.\,\ref{fig:gyield}. It is about a factor two higher than that  of
\cite{Berezinsky93}, mainly due to the higher $\pi^0$ multiplicity in
our  complete event simulations.  The gamma flux can be used to set
limits on the cosmic ray flux and the matter distribution in the
galaxy.  The experimental situation is here, however, quite different
from the case with the neutrino flux.  The gammas are detected either
directly on satellite based experiments or  with ground based air
shower arrays.  The signal rate will be higher than with neutrinos due
to the larger  interaction cross section.  The flux of gammas is,
however, very small ($\sim 10^{-4}$) compared to the  inclusive cosmic
ray flux dominated by protons. Differences in the air showers produced
by gammas and hadrons may allow access to information on the gamma
flux.

\vspace{8mm}
\noindent
{\bf Acknowledgement:} We are grateful to Ph.~Jetzer for
useful discussions.

\end{document}